\documentclass[conference]{IEEEtran}
\IEEEoverridecommandlockouts
\usepackage{cite}
\usepackage{amsmath,amssymb,amsfonts}
\usepackage{graphicx}
\usepackage{textcomp}
\usepackage{xcolor}
\usepackage{algorithm}
\usepackage{algpseudocode}

\usepackage{subfig}
\usepackage{graphicx}
\usepackage{subcaption}
\usepackage{caption}

\usepackage{booktabs}
\usepackage[table]{xcolor}
\usepackage{colortbl}

\usepackage{multirow}

\definecolor{low}{HTML}{D6EAF8}      
\definecolor{medium}{HTML}{F9E79F}   
\definecolor{high}{HTML}{F5CBA7}     

\newcommand{\scorecell}[1]{%
  \ifnum#1<2
    \cellcolor{low}#1%
  \else\ifnum#1<5
    \cellcolor{medium}#1%
  \else
    \cellcolor{high}#1%
  \fi\fi
}
\def\BibTeX{{\rm B\kern-.05em{\sc i\kern-.025em b}\kern-.08em
    T\kern-.1667em\lower.7ex\hbox{E}\kern-.125emX}}

\begin{document}

\title{NOMAD: A Multi-Agent LLM System for UML Class Diagram Generation from Natural Language Requirements\\

}

\author{\IEEEauthorblockN{Polydoros Giannouris}
\IEEEauthorblockA{\textit{Department of Computer Science} \\
\textit{University of Manchester}\\
Manchester, United Kingdom \\
polydoros.giannouris@manchester.ac.uk}
\and
\IEEEauthorblockN{Sophia Ananiadou}
\IEEEauthorblockA{\textit{Department of Computer Science} \\
\textit{University of Manchester}\\
Manchester, United Kingdom \\
sophia.ananiadou@manchester.ac.uk}

}

\maketitle

\begin{abstract}
Large Language Models (LLMs) are increasingly utilised in software engineering, yet their ability to generate structured artefacts such as UML diagrams remains underexplored. In this work we present NOMAD, a cognitively inspired, modular multi-agent framework that decomposes UML generation into a series of role-specialised subtasks. Each agent handles a distinct modelling activity, such as entity extraction, relationship classification, and diagram synthesis, mirroring the goal-directed reasoning processes of an engineer. This decomposition improves interpretability and allows for targeted verification strategies. We evaluate NOMAD through a mixed design: a large case study (Northwind) for in-depth probing and error analysis, and human-authored UML exercises for breadth and realism. NOMAD outperforms all selected baselines, while revealing persistent challenges in fine-grained attribute extraction. Building on these observations, we introduce the first systematic taxonomy of errors in LLM-generated UML diagrams, categorising structural, relationship, and semantic/logical. Finally, we examine verification as a design probe, showing its mixed effects and outlining adaptive strategies as promising directions. Together, these contributions position NOMAD as both an effective framework for UML class diagram generation and a lens onto the broader research challenges of reliable language-to-model workflows.

\end{abstract}
\begin{IEEEkeywords}
Large Language Models (LLMs), AI for Modelling, UML Diagram Generation, Model-Driven Engineering, Requirements Engineering\end{IEEEkeywords}
\section{Introduction}

Domain modelling plays a central role in software and systems development by capturing the structure of a problem domain and providing a shared conceptual understanding among stakeholders. However, constructing accurate models remains challenging, and errors introduced at this stage are costly to correct later in the development lifecycle \cite{ferreira2011reducing}. Consequently, precise domain modelling is critical for reducing downstream development effort and limiting late-stage changes.

Within Model-Driven Engineering (MDE), domain models are formalised into artefacts such as UML class diagrams. Traditional automation techniques for generating UML models from natural language rely on handcrafted, rule-based systems \cite{meng2024automated,meziane2008generating}. While interpretable, these approaches struggle to generalise due to linguistic variability and ambiguity in requirements \cite{fantechi2017ambiguity}. 
LLMs can map natural language directly to structured representations \cite{he2024llm}, and multi-agent systems have been shown to improve LLM performance by decomposing complex tasks \cite{qian2023chatdev}. However, despite UML diagram generation necessitating multiple reasoning steps, no prior work has applied this modular approach to UML diagram generation.

In this work, we present NOMAD, a cognitively inspired multi-agent system that decomposes class diagram generation into goal-directed subtasks, each handled by a specialized agent, mirroring the reasoning process of software engineers.

We evaluate NOMAD through a mixed approach: (1) a manually constructed case study based on the widely used Northwind 2.0\footnote{https://support.microsoft.com/office/northwind-2-0-developer-edition-32eb79d2-bede-4ea4-b575-0714ca8dc1e2} Microsoft Access database, which stresses scalability and enables error analysis, and (2) eight human-authored UML modelling exercises \cite{camara2023assessment}, which provide realistic natural language requirements and curated diagrams. Finally, we introduce a hierarchical error taxonomy for LLM-generated UML diagrams, which systematically categorises structural, relationship, and semantic/logical deviations. This taxonomy enables reproducible and detailed error analysis, highlighting common pitfalls in model outputs.

Our contributions are as follows \footnote{https://github.com/PolydorosG/NOMAD}:
\begin{itemize}

\item We introduce \textbf{NOMAD}, a modular, cognitively inspired multi-agent LLM framework for generating UML class diagrams from natural language requirements.

\item We construct and release a detailed \textbf{case study} based on the Northwind 2.0 database, enabling reproducible benchmarking.

\item We develop an \textbf{automatic evaluation framework} to assess the correctness of generated models, enabling scalable and automatic benchmarking.

\item We introduce the \textbf{first error taxonomy} for LLM-generated UML diagrams, which categorises deviation types, enabling systematic and reproducible error analysis.

\item We conduct a \textbf{comparative analysis} of NOMAD and single LLMs, exploring LLM self-verification to improve accuracy.

\end{itemize}

\begin{figure*}[h]
  \centering
  \includegraphics[width=0.65\linewidth]{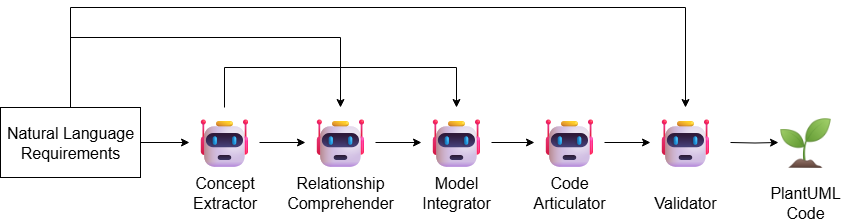} 
  \caption{The NOMAD pipeline.}
  \label{fig:pipeline}
\end{figure*}

\section{Related Work}
\subsection{Automatic Generation of UML diagrams from Requirements}

Traditional approaches to generating UML diagrams from natural language predominantly employ deterministic, rule-based systems. These pipelines typically involve a sequence of linguistic preprocessing steps such as part-of-speech (POS) tagging, dependency parsing, and ontology or heuristic-based extraction of relevant UML elements. For instance, READ \cite{bashir2021modeling} and related works \cite{more2012generating, sharma2015natural} apply handcrafted rules to map linguistic structures to UML elements. While structured and interpretable, these methods are brittle and domain-specific, often struggling with linguistic ambiguity and diverse requirement formats. More recent pipelines, such as that of Meng et al. \cite{meng2024automated}, incorporate preprocessing and sentence classification, but remain dependent on rigid grammar rules. Yang et al. \cite{yang2022towards} introduce pronoun resolution and pattern-based extraction to improve coherence, yet the core limitations persist: rule-based methods are hard to scale and adapt. Their reliance on handcrafted rules and fixed linguistic patterns limits their ability to handle the inherent ambiguity and variability of natural language. Consequently, these methods often struggle with generalising to diverse or unstructured requirements documents, motivating research into more flexible, data-driven techniques.

\subsection{Large Language Models for Modelling}
Recent research has explored the application of Large Language Models (LLMs) in software modelling tasks, particularly in generating UML diagrams. These studies either position LLMs as modelling assistants or as autonomous agents translating text to PlantUML. Fill et al. \cite{fill2023conceptual} assess the performance of GPT-4 across different types of UML diagrams and found that while it produces largely correct diagrams, they still require manual review. Similarly, Cámara et al. \cite{camara2023assessment} evaluate the syntactic correctness of PlantUML diagrams generated by LLMs and report generally accurate code. However, they identify significant limitations in semantic accuracy, particularly in larger diagrams, noting that they have serious errors with models larger than 10–12 classes. Likewise, \cite{chen2023automated} evaluate GPT-3.5 and GPT-4 with various prompt engineering techniques on a dataset of ten domain examples and find that, despite showing satisfactory domain understanding, both models are far from achieving full modelling automation. Wang et al. \cite{wang2024llms} examine the effectiveness of LLM-assisted modelling with 45 undergraduate software engineering students. Their findings suggest that while class definitions were mostly accurate, the models struggle to correctly infer relationships. Ferrari et al. \cite{ferrari2024model} investigate ChatGPT’s ability to generate sequence diagrams from structured requirements. Although the generated models generally adhere to UML standards and are reasonably understandable, issues with completeness and correctness emerge, especially when the input is contaminated with requirement smells (e.g. ambiguity). Importantly, these approaches rely exclusively on proprietary models. In contrast, NOMAD improves upon the single-agent baseline using both open-source models and closed-source models. This demonstrates that open-source alternatives within a multi-agent framework can not only match but surpass single-agent performance, enhancing both accessibility and control over the modelling process.

\subsection{Multi-Agent Systems for Software Engineering}
LLM-based multi-agent systems offer key advantages in modularity, interpretability, and system design flexibility. By distributing complex tasks across specialised agents, each guided by distinct roles or prompting strategies, these systems can manage cognitive complexity more effectively than single-agent setups \cite{li2023camel, wu2023autogen}. This modular organisation also improves transparency, as each agent’s output can be independently examined, debugged, and refined. 

For a comprehensive overview of the application of multi-agent LLM systems in software engineering, we refer readers to the study by He et al. \cite{he2024llm}, which provides an extensive survey of recent advances in this domain and serves as a close conceptual counterpart to our work.

We extend the application of these architectures to a new domain, structured modelling from natural language, where they have not yet been systematically studied.

\section{Methodology}
\label{sec:methodology}
This section presents the design of NOMAD, our cognitively inspired multi-agent framework for UML class diagram generation from natural language. The methodology focuses on the architecture and operational principles of the system and the taxonomy.

\subsection{Design Rationale}
UML class diagram generation requires heterogeneous reasoning: identifying domain entities, inferring relationships, formalising structure, and ensuring consistency. These tasks stress a single LLM’s limited context.
Instead, NOMAD adopts a decompositional design, assigning each reasoning function to a specialised agent with a structured prompt and constrained output schema, thereby reducing context interference and improving modularity.
This separation of tasks mirrors human modelling workflows and supports interpretability, controllability.

\subsection{NOMAD Pipeline}
NOMAD defines four generation agents and one verification agent, connected in a sequential pipeline (Figure \ref{fig:pipeline}). Each agent has a clearly defined role, constrained output format, and structured prompt template to ensure consistency.

\textit{\textbf{Concept Extractor}}. In NOMAD, the concept extractor identifies domain entities (classes) and their attributes from the input requirements. This way it grounds the initial model's vocabulary, starting from the most concrete and least ambiguous aspect of the specification. 

\textbf{\textit{Relationship Comprehender}}. The relationship comprehender receives the grounded vocabulary as input and draws associations, aggregations, compositions, and generalisations between the domain entities according to the system description thus, establishing the model’s semantic structure. 

\textbf{\textit{Model Integrator}}. The integrator combines entities and relationships into a structured object model in JSON, enforcing a uniform representation (JSON), thus formalising the model and removing any natural language ambiguity that might have been presented in previous outputs. This formal intermediate representation ensures composability and transparency across subsequent stages.

\textbf{\textit{Code Articulator}}. The articulator translated the previously generated representation into syntactically correct PlantUML. Despite its functionality being mostly deterministic, we find that adding an LLM agent at this step, rather than a hard-coded transformation, enables greater adaptability. The articulator can flexibly handle variations in the JSON structure, resolve minor inconsistencies, and maintain stylistic coherence in the generated PlantUML code.

\textbf{\textit{Validator}}. Finally, we include an optional verification agent that inspects the generated diagram against the original requirements. Prior work has explored probing the LLM to improve or regenerate its solution \cite{camara2023assessment}, which implicitly presumes the presence of errors. In contrast, our verifier enables the model to autonomously determine whether modifications are warranted, without presupposing mistakes or enforcing revisions functioning as a reflective reasoning step.

\subsection{Agent Coordination}

Each agent operates under contextualised inputs and strict constraints. For example, the Relationship Comprehender receives both extracted entities and requirements, while the Model Integrator must adhere to JSON schema definitions. Prompt templates define structure and formatting, ensuring outputs are composable across the pipeline. This design supports modularity: individual agents can be swapped, tuned, or extended without re-engineering the full system.

While advanced coordination mechanisms, such as inter-agent feedback \cite{yan2025beyond} or planning \cite{torreno2017cooperative}, represent promising directions, they lie beyond the scope of this work. Instead, we focus on demonstrating that modular, constrained agents are capable of generating coherent models without relying on complex orchestration.

\subsection{Verification Strategies}

In NOMAD, verification is treated as a design probe rather than a definitive solution. The current implementation includes a stand-alone verifier operating in a single-shot mode, providing an initial framework for automatically correcting the system's outputs. This setup motivates consideration of adaptive, model-aware verification strategies that can be incorporated into multi-agent workflows. Potential approaches include self-consistency \cite{kadavath2022language}, where LLMs evaluate the truthfulness of their own output, and cross-agent validation \cite{cohen2023lm}, in which multiple agents assess each other’s outputs.

By framing verification as a flexible methodological component, NOMAD supports integration of new verification modes without redesigning the entire workflow.

\subsection{Error Taxonomy for LLM-generated UML diagrams}
A core component of our methodology is a systematic framework for evaluating the quality of automatically generated UML class diagrams. While precision, recall, and F1 scores on classes, attributes, and relationships provide a useful quantitative signal, they do not capture the nature of the errors made by the system. To address this limitation, we develop a hierarchical error taxonomy that classifies deviations from reference models into coherent categories. This taxonomy enables fine-grained analysis of model quality, highlights recurring pitfalls of LLM-based generation, and provides a reproducible lens for comparing different approaches.

In this work, we use the term semantic correctness to refer to the degree to which the generated UML elements preserve the intended meaning of domain concepts and relationships as expressed in the natural language requirements, beyond mere syntactic conformity.

The taxonomy is structured around three principal dimensions of a UML class diagram: classes, attributes and relationships. Each dimension is further divided into error categories that reflect common patterns observed in our evaluations. We note that all generated diagrams were syntactically correct; therefore, this taxonomy only considers logical/semantic errors.

\begin{table*}[h]
\centering
\caption{Conceptual error taxonomy for valid UML diagrams}
\label{tab:error-taxonomy}
\begin{tabular}{p{3cm} p{4cm} p{6cm}}
\toprule
\textbf{Dimension} & \textbf{Error Type} & \textbf{Description / Example} \\
\midrule
\multirow{3}{*}{Classes} 
 & Missing & Reference class not generated \\
 & Extra & Spurious class introduced \\
 & Misrepresented & Class exists, but UML construct or role is incorrect (e.g., enumeration represented as a simple variable; class represents wrong domain concept) \\
\midrule
\multirow{3}{*}{Attributes} 
 & Missing & Generated diagram lacks one or more expected attributes \\
 & Extra & Attributes present in the generated diagram but absent in the reference \\
 & Wrong & Attribute semantics or name incorrect (e.g., identifier vs. descriptive field) \\
\midrule
\multirow{4}{*}{Relationships} 
 & Missing & Reference association, aggregation, or generalisation absent \\
 & Extra & Spurious connection between classes \\
 & Duplicate & Redundant copies of the same relationship \\
 & Misclassified & Conceptual type or direction incorrect (e.g., aggregation vs association; generalisation arrow reversed; association class represented incorrectly on a regular class) \\
\bottomrule
\end{tabular}
\end{table*}

This taxonomy enables both quantitative benchmarking and qualitative error analysis. By explicitly capturing modelling-specific deviations in classes, attributes, and relationships, it provides a reproducible framework for analysing LLM-generated UML diagrams across diverse case studies.

\section{Research Design}
\label{research_design}
This section outlines how we structure the research, the datasets we employ, the baselines against which NOMAD is compared, and the analysis methods used.

Our research design is driven by three research questions (RQs), which frame the scope of evaluation and guide both the quantitative and qualitative analyses. Each RQ corresponds to a specific aspect of UML class diagram generation: structural accuracy, error characterisation, and the role of verification.

To answer these questions, we construct two complementary datasets: (i) a controlled case study derived from the Northwind 2.0 schema, enabling fine-grained analysis, and (ii) eight supplementary, human-authored UML exercises, providing breadth and realism. We then define baselines, experimental configurations, and evaluation metrics tailored to probe the RQs. Finally, we establish analysis methods, combining quantitative measures of accuracy with a qualitative taxonomy of modelling errors.

\subsection{Research Questions}

Our research aims to answer the following research questions:

\textbf{RQ1:} \textit{Does cognitive decomposition improve UML generation compared to single-agent LLM prompting across diverse use cases?} This question examines the core methodological contribution of our work: whether decomposing the modelling task into cognitively inspired subtasks (concept extraction, relationship reasoning, model integration, and articulation) yields more accurate and consistent UML diagrams than treating the task as a single prompt.

\textbf{RQ2:} \textit{What types of modelling errors are most common in LLM-generated UML diagrams, and how can they be systematically categorised?}  
Accuracy metrics alone cannot capture the nuanced ways in which generated diagrams diverge from correct models. To provide a reproducible and interpretable account of model limitations, we construct a taxonomy of UML generation errors. This taxonomy enables us to quantify recurring modelling pitfalls across models and configurations.

\textbf{RQ3:} \textit{To what extent can verification strategies reduce errors in UML generation, and what are their limitations?} LLM-generated diagrams are prone to inconsistency and omission, motivating the need for verification. Here, we study the impact of a verifier agent on structural correctness and error reduction, and we assess the situations in which it provides meaningful improvements and those in which its effects remain limited.

\subsection{Dataset Construction}
To conduct the experiments, it was required to obtain a natural language description and a corresponding golden diagram of the described system. Although previous work has utilised such data \cite{de2024evaluating, calamo2025assessing, fill2023conceptual, wang2024llms}, we find that it is often (1) closed source, (2) omitting key UML elements for the sake of simplicity (e.g. attributes), or (3) small in size and complexity. 

To address this gap, we construct a gold-standard benchmark derived from a database schema, comprising both a reverse-engineered UML class diagram and a curated set of corresponding natural language requirements. This serves as our primary case study, enabling detailed analysis, error taxonomy development, and verification probing. To complement this evaluation and broaden coverage, we incorporate a set of eight independent human-written UML exercises drawn from prior work \cite{camara2023assessment}, allowing us to assess generalizability across diverse requirements.

\subsubsection{Case Study Construction}

For the case study, we opt for the Northwind 2.0 database, a canonical example schema originally distributed with Microsoft Access. We extract its structure, convert it to SQL, and construct a corresponding UML class diagram that faithfully represents the underlying data model.

A UML class diagram comprises classes \( C \), each with a set of attributes \( A(C) \), operations, and relationships \( R \subseteq C \times C \). As Northwind originates from a relational schema, operations are excluded from analysis due to their absence in the source model.

The conversion follows a rule-based reverse engineering process \cite{iftekhar2019reverse}:

\begin{itemize}
    \item \textbf{Tables}: Each relational table is mapped to a UML class. If a table functions as a pure join table for a many-to-many relationship, it is converted to an association between two other classes rather than as a standalone class.
    \item \textbf{Columns}: Table columns are translated directly into attributes \( a_i \in A(C) \), preserving names and types.
    \item \textbf{Relationships}: Foreign key constraints are translated into UML associations. Cardinality is inferred from key properties:
    \begin{itemize}
        \item A unique and non-nullable foreign key implies a one-to-one relationship.
        \item A non-unique, non-nullable foreign key implies a one-to-many relationship.
        \item Join tables yield many-to-many associations.
    \end{itemize}
\end{itemize}

The resulting class diagram contains 21 classes and 212 total elements, with a range of associations, aggregations, and compositions. This reference model provides a comprehensive and realistic target against which automatically generated UML diagrams can be systematically evaluated.

We implement a deterministic, rule-based transformation to generate structured natural language (NL) requirements from UML class diagrams, ensuring semantic fidelity and traceability. While the transformation is guided by formal rules, the script is authored by a human with the deliberate intent of producing requirements that read naturally and resemble human-written specifications, rather than rigid, machine-generated text.

Each class \( C \) is translated into NL requirements specifying system responsibilities to \textit{record} or \textit{maintain} its attributes \( A(C) \) as grouped semantic properties. Formally, for all classes \( C \), the requirement states: “The system shall record \( C \) information including \( A(C) \).”

Binary associations \( R_{C_i \leftrightarrow C_j} \) with cardinalities \( (m..n) \) are mapped to quantified relational constraints, e.g., “The system shall associate each instance of \( C_i \) with \( m..n \) instances of \( C_j \).”

Generalisation hierarchies \( C \triangleleft \{S_1, \ldots, S_k\} \) yield classification requirements such as “The system shall maintain instances of \( C \) classified as either \( S_1, \ldots, S_k \).”

Many-to-many relations and association classes generate requirements describing composite entities and their components (e.g., order details). Enumerated types and static classification entities \( E \) generate requirements to maintain corresponding catalogues for domain classification.

Importantly, audit-related attributes (e.g., created\_by, modified\_at), although derivable from stereotypes or shared schema conventions, are specified through separate, explicit requirements instead of being uniformly folded into the class-based rule. This reflects a practical choice to preserve clarity and modularity in the generated requirements.

Finally, structural constraints (e.g., referential integrity, naming conventions) are translated into non-functional requirements, and all generated requirements are systematically labelled using domain-specific prefixes and numeric identifiers to support traceability and standards compliance.

This approach balances formal consistency with human readability, using rules to guide the process while allowing flexibility to produce clear, natural-sounding requirements suitable for practical use.

\subsubsection{Supplementary Use Cases}
\label{sec:supplementary}

While the Northwind 2.0 database serves as our primary case study for detailed probing and verification analysis, relying on a single dataset risks overfitting our evaluation to one particular domain and requirements style. To broaden the scope and assess generalisability, we include a set of eight supplementary UML use cases drawn from prior work on UML modelling evaluation \cite{camara2023assessment}. These supplementary cases are selected because they: 
\begin{itemize}
    \item originate from domains distinct from Northwind to test domain transfer,
    \item provide human-written or curated requirements, mitigating the risk of artefacts introduced by our rule-based requirements generation process, and
    \item have been previously used in benchmarking studies, allowing direct comparison with related literature.
\end{itemize}

Each use case is originally presented as a series of conversations between human experts and ChatGPT. The experts aimed to describe the system to ChatGPT and probe its ability to replicate their goal diagram. For our purposes, we collapse each conversation into a single message that contains only the initial descriptions, omitting formatting instructions, partial solutions, and corrections provided during the interaction. The resulting descriptions, along with the expert-crafted UML diagrams, serve as additional gold standards. We evaluate NOMAD on these use cases using the same metrics as in the Northwind study (precision, recall, and F1 for classes, attributes, and relationships). This mixed evaluation design combines \emph{depth} (a single large case study with extensive probing) and \emph{breadth} (a diverse set of smaller use cases). Table \ref{tab:supplementary_usecases} reports statistics for these supplementary use cases.

\begin{table}[h]
\centering
\caption{Overview of supplementary UML use cases used for breadth evaluation. Model size is measured as the total number of UML elements (classes + relationships + attributes).}
\label{tab:supplementary_usecases}
\begin{tabular}{c c c}
\toprule
\textbf{ID} & \textbf{Model Description} & \textbf{Total Elements} \\ \midrule
UC1 & Airline–airport operations & 9 \\
UC2 & Vehicle types & 16 \\
UC3 & Car model &  17\\
UC4 & File system & 9 \\
UC5 & Robots and tasks & 17 \\ 
UC6 & Theatres and plays & 25 \\ 
UC7 & University course entities &  12\\
UC8 & Video stores & 8 \\ \bottomrule 
\end{tabular}
\end{table}

\subsection{Experimental Setup}
We generate the class diagrams for all variants by prompting GPT4o through the API interface, opting for 0 temperature to minimise randomness in the outputs and set the maximum tokens to 4096 to ensure responses are not truncated. All agents in NOMAD were implemented using the LangChain framework. We select the PlantUML textual language for UML construction, a widely used and human-readable language.

\renewcommand{\arraystretch}{1.3} 
\begin{table*}[ht]
\caption{Structural correctness results (F1 scores).}
\label{tab:rq1-results}
\centering
\begin{tabular}{llccccc}
\hline
\textbf{Model} & \textbf{Method} & \textbf{Classes} & \textbf{Attributes} & \textbf{Relationships (S)} & \textbf{Relationships (R)} & \textbf{Average} \\
\hline

\multirow{2}{*}{Small models} 
&Base              &  0.8185 &  0.7084  &  0.4617  & 0.6358 & 0.6561\\
&NOMAD             &  0.8573    &   0.7711 &  0.4955 &  0.6592 & 0.6958\\
\hline
\multirow{2}{*}{Northwind} 
               & Base   & 0.9545 & 0.5898 & 0.5211          & 0.8777       & 0.7360 \\
               & NOMAD  & 1.0000 & 0.4412 & 0.9231 & 0.9811 & 0.8360 \\

\hline
\end{tabular}
\end{table*}

\subsection{Analysis Method}

\subsubsection{RQ1}
To address \textbf{RQ1}, we propose a novel and reproducible evaluation methodology, since no standard metrics exist in prior literature for this task. Our approach directly parses PlantUML code using regular expressions to extract classes, attributes, and relationships, and employs NLTK-based normalisation \cite{bird2009natural} to handle variations in naming (e.g., plurals, capitalisation). Precision, recall, and F1-score are then computed for each element type, with relationships evaluated under both strict (“hard”) and relaxed (“soft”) matching criteria. This methodology enables a rigorous, automated, and interpretable assessment of LLM-generated UML diagrams. The following formal definitions specify how each metric is computed.

Let \( C_g \) and \( C_m \) denote the sets of classes in the gold (reference) and LLM-generated UML class diagrams, respectively. For each class \( c \in C_g \), let \( A_g(c) \) be the set of attributes in the gold diagram, and similarly \( A_m(c) \) for \( c \in C_m \) in the model diagram. Relationships are defined as tuples \((c_s, c_t, t_r)\), where \( c_s \) and \( c_t \) are source and target classes, and \( t_r \) is the relationship type. The sets of relationships in the gold and model diagrams are \( R_g \subseteq C_g \times C_g \times T \) and \( R_m \subseteq C_m \times C_m \times T \), respectively.

Classes are matched by exact name equality, i.e., the matched classes are \( C_g \cap C_m \). For these matched classes, attributes are matched by exact string equality: 
\[
\bigcup_{c \in C_g \cap C_m} A_g(c) \cap A_m(c).
\]

Relationships are evaluated under two criteria. The \emph{hard} matching requires exact agreement on source class, target class, and relationship type:
\[
R_g \cap R_m,
\]
while the \emph{soft} matching only requires that the source and target classes are connected, regardless of relationship type and directionality:
\[\big\{ (c_s, c_t, t_r) \in R_g \mid \exists t_r', \; (c_s, c_t, t_r') \in R_m \text{ or } (c_t, c_s, t_r') \in R_m \big\}.
\]

For each element type (classes, attributes, and relationships), we denote the sets in the gold and model diagrams as \(E_g\) and \(E_m\), respectively, and their intersection as the matched elements \(E_\cap\). Precision and recall are defined as
\[
\text{Precision} = \frac{|E_\cap|}{|E_m|}, \quad
\text{Recall} = \frac{|E_\cap|}{|E_g|},
\]
with the F1-score computed as the harmonic mean of precision and recall:
\[
\text{F1} = \frac{2 \cdot \text{Precision} \cdot \text{Recall}}{\text{Precision} + \text{Recall}}.
\]

\subsubsection{RQ2}
For \textbf{RQ2} we conduct a qualitative analysis of all the LLM outputs and find that most errors fall under certain categories. To this end, we create the first taxonomy of UML generation errors by LLMs. The taxonomy was initially developed in an ad-hoc manner based on known UML modelling patterns and anticipated LLM errors. It is intentionally flexible and designed to accommodate empirical observations: categories with no observed errors remain in the hierarchy to ensure completeness, while new error types discovered during annotation were incorporated iteratively. This hybrid approach ensures both coverage of anticipated mistakes and adaptability to real-world LLM outputs, providing a robust framework for systematic evaluation. We utilise this taxonomy to analyse error types and distributions in the case study and compare error types that are mitigated or exacerbated by the use of NOMAD. 

\subsubsection{RQ3}
To address \textbf{RQ3}, we examine the impact of the Validator agent on the quality of UML diagrams generated within the NOMAD system. Our analysis focuses on quantitative changes in diagram accuracy, comparing F1-scores between outputs produced with and without the Verifier. To further assess the generality of the verification effect, we repeat this comparison using both GPT-4o and DeepSeek-V3 as base LLMs, allowing us to determine whether self-verification yields consistent improvements across different model architectures.

\section{Results \& Discussion}
This section presents results and analyses structured around the three research questions (RQs) introduced in Section \ref{research_design}.

\subsection{RQ1}
RQ1 investigates whether decomposing the UML generation process into specialised agents improves structural accuracy compared to a single-agent baseline.

To answer this, we evaluate structural fidelity using F1 scores over classes, attributes, and relationships (strict and relaxed matching). Results are reported separately, for the breadth study on eight supplementary use cases and the in-depth Northwind case study. This design allows us to assess both generalisation across diverse requirements and detailed behaviour on a larger schema.

\subsubsection{Breadth Evaluation}
Table~\ref{tab:rq1-results} shows results for small models on the eight supplementary use cases. The base single-agent model achieves reasonable class F1 (0.82) and attribute F1 (0.71), but struggles with relationships under strict matching (0.46) and performs moderately under relaxed matching (0.64). Introducing NOMAD improves performance across all dimensions: class F1 rises to 0.86, attribute F1 to 0.77, strict relationship F1 to 0.50, and relaxed relationship F1 to 0.66. Overall F1 increases by 4\% from 0.66 to 0.70. These results confirm that decomposing the modelling task into subtasks yields systematic improvements across a variety of domains.


\subsubsection{Depth Evaluation}
Table~\ref{tab:rq1-results} reports detailed results for the Northwind case study, comparing GPT-4o under both direct prompting (Base) and the NOMAD multi-agent framework. Several trends emerge.

\textbf{Classes.} GPT-4o already performs strongly at class identification when prompted directly (0.95). NOMAD raises this score to complete recovery (1.00), suggesting that high-level entity recognition is a relatively solved capability for current LLMs.

\textbf{Attributes.} Attribute extraction remains the weakest area. GPT-4 achieves 0.59 under direct prompting and slightly worsens to 0.44 with NOMAD. This drop suggests that while NOMAD improves structural reasoning, its decomposition introduces stricter constraints that can suppress attribute recall. The finding reinforces attributes as an open research challenge requiring additional strategies beyond decomposition.

\textbf{Relationships.} Relationship modelling is where NOMAD provides the largest gains. GPT-4 jumps from 0.52 (strict F1) under direct prompting to 0.92 with NOMAD. Under relaxed matching, the model surpasses 0.90. These improvements show that NOMAD’s decomposition helps the model capture inter-entity dependencies more consistently, reducing both omissions and spurious links.

\textbf{Overall.} Averaged across all categories, GPT-4o benefits from NOMAD (0.74 $\rightarrow$ 0.84). While class identification is nearly perfect, attribute extraction remains challenging. Attributes are often implied rather than explicitly stated, and requirements may describe functionality rather than concrete data fields. Moreover, attributes tend to be highly domain-specific (e.g., audit fields versus business logic), underscoring a broader research challenge rather than a limitation of any single approach.

\subsubsection{Summary}
Across both breadth and depth settings, NOMAD consistently improves structural accuracy, with the largest gains observed in relationship modelling. Class detection is largely solved across models, whereas attribute extraction and relationships remain the weakest subtasks. Interestingly, despite Northwind being a much larger case study (21 classes and 212 total elements), its scores are generally higher than those of the eight supplementary use cases. We attribute this to the post hoc construction of Northwind requirements, which were well formatted and internally consistent, in contrast to the fuzzier, human-authored descriptions in the supplementary set. This discrepancy underscores two important points: first, that LLMs are sensitive to redundancy and ambiguity in natural language, and second, that reliable UML generation depends not only on modelling methodology but also on the quality of the requirements themselves, illustrating a “garbage in, garbage out” dynamic in language-to-model workflows.

\begin{table*}[ht]
\caption{Structural correctness results with and without the addition of the verifier agent.}
\label{tab:rq3:quant}
\centering
\begin{tabular}{llccccc}
\hline
\textbf{Model} & \textbf{Method} & \textbf{Classes} & \textbf{Attributes} & \textbf{Relationships (S)} & \textbf{Relationships (R)} & \textbf{Average} \\
\hline

\multirow{2}{*}{GPT-4o} 
& NOMAD        & 1.0000 & 0.4412 & 0.9231 & 0.9811 & 0.8360 \\
& NOMAD$_v$& 1.0000 & 0.618 & 0.9231  & 1.0000  & 0.8850 \\
\hline
\multirow{2}{*}{DeepSeek V3} 
               & NOMAD     & 0.9545 & 0.4130 & 0.6429 & 0.9123 & 0.7310 \\
              & NOMAD$_v$   & 0.9545 & 0.6859 & 0.7809 & 0.8895 & 0.8280 \\

\hline
\end{tabular}
\end{table*}

\subsection{RQ2}
RQ2 explores what types of modelling errors occur most frequently in LLM-generated UML diagrams and how they can be systematically categorised.

\begin{figure}[h]
  \centering
  \includegraphics[width=1\linewidth]{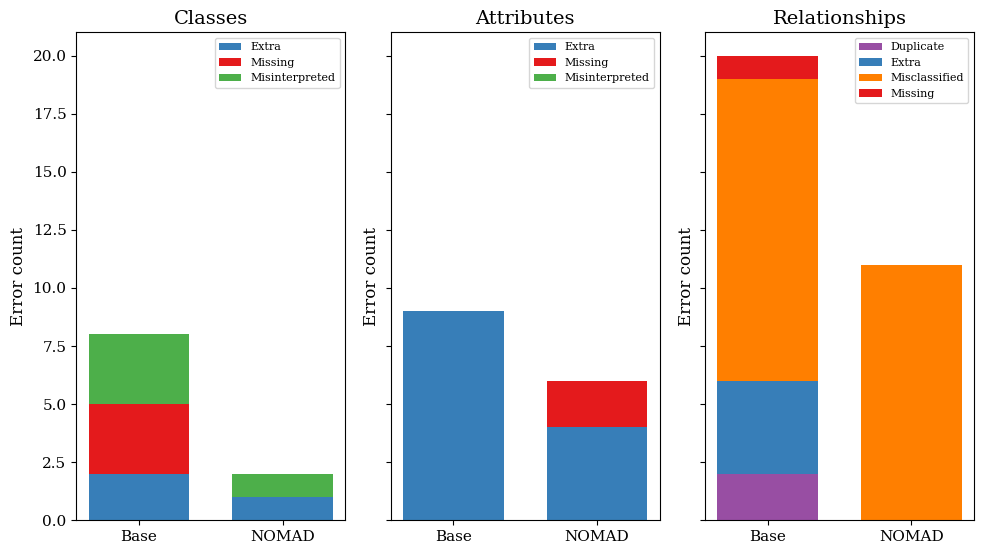} 
  \caption{Comparison of error types in baseline vs NOMAD UML outputs. Each stacked bar shows the count of errors per dimension, categorised by the proposed taxonomy.}
  \label{fig:bar8}
\end{figure}

To study this, we annotate each of the diagrams using the proposed taxonomy. We aggregate the scores and present the errors across the three UML dimensions (classes, attributes, and relationships) for both the baseline and NOMAD outputs over the 8 diagrams in Figure \ref{fig:bar8}. 

The chart illustrates several key trends. For classes, the baseline exhibits multiple extra, missing, and misinterpreted errors, whereas NOMAD eliminates all missing classes and reduces misinterpretations and extraneous classes. For attributes, NOMAD reduces the number of spurious attributes by more than half compared to the baseline, though it tends to miss some that were otherwise included. It should be noted that we do not count missing or extra attributes for classes that are not included in the golden diagram to avoid over-penalising class errors. For relationships, baseline outputs are dominated by misclassified connections, along with some extra, duplicate, and missing relationships. NOMAD eliminates all missing, extra, and duplicate relationship errors, although a significant degree of misclassification remains, mostly due to incorrect directionality or relationship type.

Applying the same taxonomy to the Northwind case study produces a closely aligned pattern. For \textit{classes}, the baseline introduces two extraneous elements, while NOMAD generates a perfectly aligned set with no missing or extra classes. For \textit{attributes}, both approaches show a trade-off between completeness and precision: NOMAD slightly increases missing and extra attributes (10 and 9, respectively) but reduces misinterpretations from 3 to 1, indicating more semantically coherent naming. The most pronounced difference appears in \textit{relationships}, where the baseline produces 22 misclassified connections, compared to only 3 under NOMAD, with all other error categories nearly eliminated. Overall, these Northwind results reinforce the trends observed across the supplementary cases, NOMAD consistently removes structural noise and improves semantic fidelity, while attribute coverage remains a nuanced challenge.

Overall, the chart highlights that NOMAD substantially reduces high-impact structural errors, particularly in classes and relationships, while attribute errors show a modest trade-off between completeness and precision. This visualisation makes it clear which error types are most effectively mitigated by NOMAD and which remain challenging, providing a nuanced, taxonomy-driven perspective on LLM-generated UML outputs.

\subsection{RQ3}
RQ3 examines the role of verification in improving model correctness and consistency. 
 
To this end, we evaluate the impact of the verifier agent on the Northwind case study by measuring F1-scores across structural elements—classes, attributes, and relationships. The verifier agent receives the final UML code generated by the modelling pipeline, together with the original natural language description and is prompted to identify and correct any inconsistencies. To ensure generalizability, experiments are conducted using both GPT-4o and DeepSeek V3 as the LLM backbone. Table \ref{tab:rq3:quant} summarises the quantitative results.

Across both models, the addition of the verifier agent improves overall structural correctness, with particularly notable gains in attribute and relationship quality. For DeepSeek V3, the class structure remains unchanged, suggesting that the base NOMAD pipeline already identifies core entities to the best of the model's abilities. However, attribute correctness increases substantially by more than 27.3\%, demonstrating the verifier’s effectiveness in recovering omitted or misclassified attributes. Relationship performance also improves, rising from 0.6429 to 0.7809 under strict matching and slightly decreasing under relaxed matching (0.9123 → 0.8895). This indicates that the verifier primarily enhances relationship directionality and type accuracy rather than basic connectivity.

For GPT-4o, where baseline structural accuracy is already high, the verifier still yields measurable gains. Attribute F1 increases from 0.4412 to 0.618, and relationship directionality achieves full correctness. The model already produces perfect class identification and near-perfect relationship semantics, yet the verifier further enhances overall consistency, raising the average F1 from 0.8360 to 0.8850. Qualitatively, this improvement manifests in resolving subtle redundancies (e.g., duplicate associations), refining attribute coverage in inherited classes, and correcting occasional directionality mismatches.

While these results demonstrate consistent improvements across both LLM backbones—as reflected in the aggregate F1 gains reported in Table \ref{tab:rq3:quant}, our current evaluation remains coarse-grained. Table \ref{tab:rq3:quant} quantifies the verifier’s overall contribution, but it does not isolate which specific error types the verifier corrected or introduced. In particular, we do not yet measure performance at the level of individual error categories such as directionality, cardinality, attribute naming, or broader semantic versus syntactic repairs. As a result, the verifier’s repair precision and its potential side-effects cannot be fully characterised from the present results alone. A fine-grained analysis of verifier behaviour will be pursued in future work to substantiate the verifier’s targeted effectiveness and validate semantic verification in multi-agent modelling pipelines.

In summary, the verifier agent plays a complementary role in both systems: while the base NOMAD framework establishes a structurally coherent baseline, the verifier enhances fine-grained consistency and completeness. These results suggest that introducing post-hoc structural verification can reliably elevate the fidelity of model-based generation pipelines without compromising their core outputs. 

\subsection{Qualitative Error Analysis and Causal Interpretation}
While quantitative metrics show consistent improvement across datasets, a qualitative inspection of NOMAD’s outputs helps explain its remaining limitations and the causal mechanisms behind them. The analysis below highlights representative error cases and discusses how decomposition affects both model reliability and semantic alignment.
\begin{table*}[h]
\centering
\begin{tabular}{p{3cm}p{4cm}p{3cm}p{4cm}}
\toprule
\textbf{Example} & \textbf{Description} & \textbf{Error Type} & \textbf{Likely Cause} \\
\midrule
Robot type modeling & Requirement: “robots can be of type ...”. NOMAD modeled an \texttt{enum + attribute}, while the gold standard used inheritance. & Structural ambiguity & Ambiguity in natural language: “type” interpreted as property rather than subclass cue. Penalized under strict matching. \\
Inter-agent attribute removal & Concept extractor assigned attributes to both classes A and B, but the relationship comprehender later inferred inheritance, making one attribute set redundant. & Inter-agent inconsistency & Sequential communication loss between agents; relationship updates not fully propagated backward. \\
Persistent class identification errors & Both NOMAD and single-agent models occasionally miss or misclassify classes. Identifying entities remains easier than producing a coherent UML diagram, but LLMs are still imperfect at consistent schema induction. & Shared modelling limitation & Inherent difficulty for LLMs in maintaining stable conceptual boundaries and type consistency across generated artifacts. \\
\bottomrule
\end{tabular}
\caption{Representative NOMAD error cases and their causal factors.}
\label{tab:error_examples}
\end{table*}

The cases presented in Table \ref{tab:error_examples} reveal three main sources of error. First, some limitations stem from the underlying language models themselves: even though class identification is simpler than full UML synthesis, LLMs are not yet reliable at maintaining consistent conceptual boundaries, leading to persistent misclassifications shared by both single- and multi-agent approaches. Second, NOMAD occasionally produces semantically reasonable but structurally divergent designs, such as modelling “robot types” using enumeration rather than inheritance. These variants reflect ambiguity in the natural-language specification rather than incorrect reasoning, yet they are penalised under our structural evaluation criteria. Third, inter-agent inconsistencies arise from sequential information loss: later agents, such as the relationship comprehender, may override or contradict earlier inferences made by the concept extractor, particularly when new relationships invalidate previously assigned attributes. For example, in the case of vehicles, the concept extractor may assign attributes like \texttt{position}, \texttt{speed}, and \texttt{environment} to \texttt{Vehicle}, \texttt{LandVehicle}, and \texttt{MarineVehicle} individually. Once the relationship comprehender infers that \texttt{LandVehicle} and \texttt{MarineVehicle} inherit from \texttt{Vehicle}, some attribute assignments become redundant or inconsistent across the hierarchy. Finally, we observe that the verifier agent can correct some of these “surface-level” issues, such as duplicated attributes or incorrect relationship directionalities. However, its effect is limited to postprocessing refinements; it does not fundamentally restructure the diagram, as allowing it to completely overhaul the hierarchy would defeat its role as a verification agent rather than a generative one. 

Decomposition appears to help as the complexity and size of the requirement description increase. Larger and more detailed inputs benefit from specialised agents that can focus on distinct reasoning subtasks, resulting in improved scalability and structural coherence. However, for smaller or simpler diagrams, decomposition can introduce unnecessary fragmentation and even degrade performance, as the overhead of coordination outweighs the benefits of specialisation. Furthermore, the sequential topology of agents can lead to communication losses when relational decisions clash with earlier attribute assignments. Overall, these findings suggest that while decomposition supports modular reasoning, limited bidirectional feedback or shared contextual memory could mitigate such inconsistencies and enhance performance across varying diagram complexities.

\section{Conclusion}
This paper introduced NOMAD, a cognitively inspired multi-agent framework for generating UML class diagrams from natural language requirements. Unlike monolithic prompting approaches, NOMAD decomposes the modelling process into specialised, role-driven agents that emulate the reasoning steps of a human modeller—from concept extraction to verification. 

Through a mixed evaluation design that combines the Northwind 2.0 case study with a suite of human-authored UML exercises, NOMAD demonstrated consistent gains in structural accuracy and semantic fidelity compared to single-agent baselines. The framework achieved near-perfect class identification and substantial improvements in relationship reasoning, confirming that multi-agent cognitive decomposition enhances the robustness and coherence of LLM-generated models. At the same time, the study revealed that fine-grained attribute extraction remains a persistent limitation, reflecting the inherent ambiguity and variability in natural language specifications rather than model design alone. This underscores that, while multi-agent and semantically decomposed approaches offer clear benefits, they also consistently introduce errors and require careful design, monitoring, and post-hoc verification.

We further introduced the first error taxonomy for LLM-generated UML diagrams, offering a reproducible and interpretable structure for analysing deviations across class, attribute, and relationship dimensions. This taxonomy enabled a deeper understanding of how modular decomposition mitigates specific error types while exposing those that remain challenging. Finally, by integrating an LLM-based verifier, we demonstrated that post-hoc verification can meaningfully improve model completeness and consistency without compromising structural soundness.

Overall, this work positions NOMAD as both an effective method for UML synthesis and a conceptual lens through which to explore reliable language-to-model transformation. By combining cognitive decomposition with systematic evaluation, NOMAD contributes toward a more interpretable, verifiable, and extensible paradigm for AI-assisted modelling within Model-Driven Engineering.

\section{Future Work}
Future work will extend the NOMAD architecture to support a broader range of modelling formalisms, including behavioural and hybrid diagrams (e.g., SysML, sequence diagrams). We also plan to integrate human-in-the-loop feedback, explore reinforcement learning and retrieval-based strategies for adaptive orchestration, and investigate multimodal grounding for richer model synthesis. More broadly, we propose multi-agent decomposition as a principled and modular approach to advancing trustworthy, traceable, and extensible AI-assisted modelling in MDE.

\bibliographystyle{IEEEtran}
\bibliography{refs}

\end{document}